\documentclass[AMA,LATO1COL]{WileyNJD-v2}

\articletype{Review Article}%

\usepackage{graphicx}
\usepackage{textcomp}
\usepackage{xcolor}
\usepackage{comment}
\usepackage{url}
\usepackage{verbatim}
\usepackage[utf8]{inputenc} 
\usepackage[T1]{fontenc}
\usepackage{hyperref}
\usepackage{float}

\newcommand{\blue}[1]{{\color{blue} #1}}

\begin{document}

\title{Can Existing Approaches Manage Dynamic and Large Business Processes enacted through Systems-of-Systems?}

\author[1]{Maria Istela Cagnin}

\author[2]{Elisa Yumi Nakagawa}

\authormark{CAGNIN AND NAKAGAWA}

\address[1]{\orgdiv{College of Computing}, \orgname{Federal University of Mato Grosso do Sul}, \orgaddress{\state{Campo Grande}, \country{Brazil}}}

\address[2]{\orgdiv{Department of Computer Systems}, \orgname{University of São Paulo}, \orgaddress{\state{São Carlos}, \country{Brazil}}}

\corres{Maria Istela Cagnin. \email{istela.machado@ufms.br}}


\abstract{
In the era of joint ventures (JV) and mergers \& acquisitions (M\&A), dynamic and large business processes can emerge to achieve broader business goals and are often formed from business processes of distinct organizations. Software systems of such distinct organizations should support these larger processes and, for this, they need to communicate among them forming the so called Systems-of-Systems (SoS). 
However, the management of these larger processes and the correspondent SoS has been currently a complicated challenge for the alliances of organizations.
In this scenario, the main contribution of this paper is to discover \textit{what} has been proposed in the literature to manage these processes. 
We analyzed possibly all existing approaches and 
the findings point out that many of them cannot provide understanding of the whole large processes and, more importantly, they do not address completely the SoS inherent characteristics. 
We also highlight the next 
research directions to make possible the management of these dynamic, complex, interconnected business processes that have increasingly crossed several critical domains.
}


\maketitle

\section{Introduction}\label{sec:introduction}

In the current scenario of alliances of organizations (JV and M\&A), each organization inevitably presents different business processes and respective goals, which could be 
joined to reach broader business goals. These goals are only tangible when the organizations' business processes properly interplay forming a large business process \cite{Cagnin2021}. In turn, business processes are essential to understand how organizations work \cite{weske2019} and they represent a collection of interrelated activities executed by one or more organizations working together to achieve a common business purpose \cite{ko2009}. 
During the business process life cycle, diverse activities (modeling, analysis, improvement, enactment and monitoring) must be carry out as they are essential for a holistic Business Process Management (BPM) \cite{weske2019,van2013}.

The business processes of each organization are sometimes supported by software systems, which, in the context of 
an alliance of organizations, should to some extent communicate and interoperate among them forming complex, large software-intensive systems referred to as Systems-of-Systems (SoS)~\cite{Dersin14SSYS, Maier1998}. 
Increasingly present in our daily lives and supporting several critical sectors of the society, e.g., transportation, health, telecommunication, and smart cities, just to mention some, SoS can support the accomplishment of larger business goals (or SoS missions, i.e., system activities to reach stakeholders goals \cite{Beale2006}) that cannot be addressed by any of their constituent systems working alone. For instance, in the case of the Brazilian public health system, a business goal is \textit{``Reducing patient wait time for a hospitalization''}. To entirely accomplish this goal, business processes from different organizations involved (e.g., mobile emergency care services responsible for pre-hospital emergency, mobile care units that refer to the ambulances, and hospitals) must interplay and take part of a dynamic large process, aiming at reaching that business goal. However, the coordination of so different business processes, public organizations with different natures, and also a resulting interconnected, large process and the corresponding software systems is not a trivial task. Besides that, these processes supported by SoS have challenges: (i) identification of the correct subset of elements of each involved process that effectively help to fulfill the alliance's business goals; (ii) additional complexity due to multiple and concomitant process life cycles; (iii) processes owned and managed by independent, distinct organizations and, consequently, limitations on the interplay and exchange of information among them exist; (iv) consistency needs among the involved processes and conformance needs between the large process and the SoS; and (v) guarantee the key properties of the SoS being properly addressed at the process level. All these peculiarities make these large business processes differ from traditional business processes; and hence, the traditional approaches for managing business processes could not work. 

The main contribution of this paper is to 
investigate \textit{what} has been proposed in the literature to manage large business processes of alliances of organizations
that count on SoS
and are referred herein as Systems-of-Systems Business Processes (SoS BP). This
paper specifically scopes the intersection between BP and SoS, as business aspects are essential to comprehend SoS \cite{Holt2012} and, at the same time, SoS is a hot topic requiring further research to advance both its state of the art and practice. For this, we 
searched for possibly all existing approaches and deeply analyzed them based on the most current and relevant literature regarding dynamic business process \cite{van2013,Weber2009, Fahland2009, Lam2015, Haisjackl2016, rodriguez2018b}, BPM \cite{weske2019, Szelagowski2019b}, and networked enterprises  (e.g., extended enterprise, virtual enterprise, and so forth) \cite{gou2000,gou2003,meng2006,baina2006,vanderhaeghen2007,li2010,akatkin2019}.

As the main result, we found there are different initiatives to mainly model SoS BP; but they do not provide means to better understand the large processes as a whole together with their business goals. They do not cover other activities (analysis, improvement, enactment and monitoring) and also do not fully address the SoS intrinsic characteristics, including the dynamism. Important research directions need to be still consolidated; hence, we also point out the most urgent ones.

The remainder of this paper is organized as follows. Section~\ref{sec:back} presents background on BPM and SoS. Section~\ref{sec:reporting} briefly describes the research method and reports the results in details.
Section~\ref{sec:discussions} discusses these results
and perspectives for future research. Finally, Section~\ref{sec:conclusion} concludes this work.

\vspace{-0.3cm}
\section{Background} \label{sec:back}



BPM is a body of principles, methods, and tools to discover, analyze, redesign, execute, and monitor business processes of organizations \cite{dumas2013}, making them more efficient and competitive. 
A \textit{business process} involves several elements, including events (that trigger the execution of one or more activities), decision points, actors, resources, and outcome that bring value to organizations \cite{dumas2018}. 
Besides, \textit{dynamic business process} is characterized by three major requirements \cite{Weber2009}: (i) support for \textit{flexibility} (process models are customized at design-time or run-time based on constraints); (ii) support for \textit{adaptation} (processes cope with expected and unexpected exceptions); and (iii) support for \textit{evolution} (process models change when the business process evolves). In turn, \textit{business process model} is an tangible artifact to express the business often in a graphical way and, by capturing several paths in which the business can go through, this model is the core artifact during the entire process life cycle \cite{van2013,dumas2018}.
This model also provides an understanding about the business operation and shares knowledge about the organization
\cite{van2013, dumas2018}. Business process model is obtained through business process modeling that embraces  \cite{van2013}: (i) identification of relevant processes and their relationships; and (ii) specification  of processes in their current state (as-is process model). 
Besides modeling, there are also other activities denominated with different terms in 
literature.
In this paper, we adopted modeling, analysis, improvement, enactment, and monitoring, all 
performed in a continuous way during the process life cycle and primordial for an adequate BPM \cite{weske2019,van2013}. In a nutshell, \textit{modeling} then creates design-time business models. \textit{Analysis} examines the models at design-time and/or run-time to detect syntactic, structural, and behavioral issues. \textit{Improvement} identifies changes in the process models to address issues identified in the previous activity, 
aiming to obtain to-be process models. \textit{Enactment} deals with to-be process models at run-time to support the operation of organizations' businesses and it involves organizational change management (refers to changes in how all participants involved in the process work) and process automation (refers to the development and deployment of software systems that support the process). Finally, \textit{monitoring} collects data about the to-be processes at run-time 
\cite{weske2019,van2013}.

SoS refer to a class of intensive-software systems that present a set of five particular characteristics, namely, emergent behavior, evolutionary development, geographic distribution, operational independence, and managerial independence \cite{Maier1998}. Initially defined from the Systems Engineering perspective, these characteristics have 
remained consensual, even considering many variants proposed by other authors \cite{Nielsen2015,baldwin2009}. Three characteristics are directly related to the nature of constituent systems: (i) \textit{operational independence} (each constituent operates independently, having its own
mission and resources); (ii) \textit{managerial independence} (constituents may present independent management and evolve in ways not foreseen when they originally joined to particular SoS); and (iii) \textit{geographic distribution} (constituents are geographically or virtually distributed). 
Considering the SoS as a whole, the two characteristics are: 
(i) \textit{emergent behavior} (new behaviors that a constituent alone is not able to deliver emerge when constituents interact and cooperate among them to achieve SoS missions); and
(ii) \textit{evolutionary development} (constituents continually evolve by their own, implying evolution in the SoS; besides, SoS also evolve due to changes in their environment and, 
as a consequence, they present dynamic architecture, requiring then new organization of their constituents). 
Finally, the combination of all these
characteristics turns SoS naturally dynamic at run-time 
concerning their architecture, constituents, environment, and missions.


\vspace{-0.3cm}
\section{Research Method and Results}
\label{sec:reporting}

To conduct this work, we carried out a comprehensive literature review following the systematic review (SR) process \cite{Kitchenham2007}, which is composed of three phases (planning, conduction, and reporting). Section \ref{sec:methodology} presents the main elements of the planning phase that are the research questions as well as a summary of the conduction phase. 
While Section~\ref{sec:overview} presents an overview of the primary studies selected, Sections \ref{sec:discussion_rq1} to \ref{sec:discussion_rq5} detail the result of our SR.

\vspace{-0.3cm}
\subsection{Planning and Conduction}
\label{sec:methodology}

%
%
%
For the \textbf{\underline{planning}}, five research questions (RQ) were defined:

\begin{list}{\labelitemi}{\leftmargin=0.5em}

    \item \textit{RQ1 - Which are the rationales (needs) to deal with the SoS BP?}
   (\textit{Rationale:} To find the motivations that leaded to the proposal of approaches to deal with SoS BP and identify the phases of SoS life cycle 
   in which SoS BP can aid.)
   
    \item \textit{RQ2: How have the SoS BP been modeled?}
    (\textit{Rationale:} Modeling is a core activity in BPM; hence, SoS BP should be also adequately modelled.)
   
    \item \textit{RQ3: Which SoS characteristics have been considered in the SoS BP modeling?}
    (\textit{Rationale:} To verify if there is a concern 
    with the SoS characteristics 
    during 
    SoS BP modeling, as such characteristics directly impact the behavior of SoS BP.)
    
    \item \textit{RQ4. Which activities of the SoS BP life cycle have been coverage by the studies?}
    (\textit{Rationale}: To effectively manage SoS BP, 
    modeling, analysis, improvement, enactment, and monitoring should be performed.)
    
    \item \textit{RQ5: Which is the maturity level of each approach found?}
    (\textit{Rationale:} The more mature the approaches are, the more likely they are to be adopted by organizations.)
\end{list}

The strategy for searching studies to answer the aforementioned RQ was based on two main keywords, namely ``\textit{business process}'' and ``\textit{System-of-Systems}''.
Additional terms were also identified together with the opinion of experts in business process and SoS. After 
various refinement iterations and string calibration, the final string was as follows: \textit{(``business process'' OR ``business model*'') AND (``System-of-Systems'' OR ``System of Systems'' OR ``SoS'' OR ``Systems-of-Systems'' OR ``Systems of Systems'')}.
Following, we decided to perform an automatic search on Scopus\footnote{\url{https://www.scopus.com}}, because it is an abstract and citation database that covers studies published in more than 10,000 publishers, including the main ones in BPM and SoS fields. After that and intending to assure the finding of possibly all existing studies, we also included the backward snowballing \cite{Wohlin2014} to possibly identify new studies.
Selection criteria (i.e., inclusion criteria (IC) and exclusion criteria (EC)) allow to include relevant studies, and to exclude non-relevant ones:

\begin{itemize}
    \item IC1: Study presents how to address business process in the SoS context or discusses about it. 
    \item EC1: Study does not address business process and SoS.
    \item EC2: Study is an editorial, keynote, panel discussion, technical report, table of contents, short course description, or summary of a conference/workshop.
    \item EC3: Study is written in a language other than English.
    \item EC4: Study is a previous version of a more complete one on the same research, of the same authors.
    \item EC5: The full text of the study is not available.
\end{itemize}

The following threats to validity of this work were identified: 

\begin{itemize}
\item \textit{Missing of important studies:} the search for studies in our SR used only one publication database, but it indexes other important digital libraries, such as ACM and IEEE, and a considerable number of more than 10,000 publishers. To be as inclusive as possible, no limit on publication date was placed. Aiming not missing important studies, we also applied snowballing from the reference list of each study. We also performed extra searches using ``cyber-physical system'', ``large scale system'', ``dynamic system'', ``complex system'', ``distributed system'', ``internet of things'', ``cloud computing'', ``cloud of things'', ``self-adaptive system'', ``autonomous system'', ``systems-of-information systems'', ``smart* system'' combined with ``business process'' and ``model*'' to cover all possible studies of other kind of systems that could be used in SoS BP. Results obtained with these extra searches were related to modeling language extensions for representing interaction of physical resources in business processes, distribution of business process fragments on the cloud (i.e., decentralized execution of business process), business process outsourcing on the cloud, and extraction of implicit goals (``which results must be addressed'' instead of ``how things must be executed'') from BPMN workflow description to support dynamic workflow systems. 
Moreover, we did not regard the keyword ``dynamic business process'' as the existing studies usually address the dynamism of a unique process without encompassing large processes, which is the interest of our work. Therefore, we believe possibly all relevant studies on SoS BP were considered in this work;

\item \textit{Reliability of review:} to assure an unbiased studies selection and reliability of our work, we rigorously followed the steps and phases of the SR process, including the definition of the SR protocol. We also concern in providing in this document enough details of this protocol to make possible the reproducibility of this SR. Moreover, SoS and business process experts supported the entire selection process and all disagreements were solved through discussion. Hence, we can ensure the way this SR was conducted; and 

\item \textit{Data extraction:} another threat referred to how the data was extracted from the studies, since not all information was obvious to be extracted and some information had to be interpreted. To ensure the validity of our results, discussions with experts were carried out.
\end{itemize}

We adopted Mendeley\footnote{\url{https://www.mendeley.com}} and MS Excel to support the SR conduction. A data extraction form recorded relevant information from each selected studies that was discussed and reviewed in consensus meetings. 


During the \textbf{\underline{conduction}} from October to December, 2019, primary studies were identified, selected, and evaluated using the selection criteria. 
We obtained a total of 106 studies from Scopus. Following, the title, abstract, and when necessary, the introduction and conclusion sections were read and the selection criteria were applied. As result, a total of 29 studies were included. The full text of these studies was then read and the selection criteria were again applied. As a result, 11 studies were selected for data extraction. 
Immediately after, we carried out the snowballing looking for studies cited by the 11 selected studies. This step required to additionally check 293 
studies, but none relevant for our SR. 
We are convinced that the small amount of studies is due to \blue{the fact that} the research in the intersection between SoS and BP is still very incipient.
Table~\ref{tab:primarystudies} presents the 11 studies, their ID, 
title, venue, venue type (J for 
journals or E for 
events (conferences and workshops)), authors, and publication year.

\begin{table*}[h]
\footnotesize
\begin{center}
\caption{Final list of selected primary studies to data extraction.}
\label{tab:primarystudies}
\begin{tabular}{|p{0.35cm}|p{6.5cm}|p{4.5cm}|p{1.0cm}|p{1.5cm}|p{0.5cm}|p{0.5cm}|}
\hline \textbf{ID} & \textbf{Title} & \textbf{Venue} & \textbf{Venue Type} & \textbf{Authors} & \textbf{Pub. Year }& \textbf{Ref.} \tabularnewline
\hline S1 & A Contribution of System Theory to Sustainable Enterprise  Interoperability Science Base & Computers in Industry & J & Ducq et al. & 2012 & \cite{Ducq2012}  \tabularnewline
\hline S2 & Business Interactions Modeling for Systems of Systems Engineering: Smart Grid Example & International Conference on System of Systems Engineering & E & Arnautovic et al. & 2012 & \cite{Arnautovic2012}   \tabularnewline 
\hline S3 & An Iterative and Recursive Model-Based System of Systems Engineering (MBSoSE) Approach for Product Development in the Medical Device domain & INCOSE Italia Conference on Systems Engineering & E &  Ciencia & 2016 & \cite{Ciancia2016}  \tabularnewline
\hline S4 &  Contributing to the GEO Model Web Implementation: A Brokering Service for Business Processes & Environmental Modelling \& Software & J & Santoro et al. & 2016 & \cite{Santoro2016}  \tabularnewline 
\hline S5 & Pattern-Based Engineering of Systems-of-Systems for Process Execution Support & International Conference on Human-Computer Interaction & E & Fleischmann et al. & 2016 & \cite{Fleischmann2016} \tabularnewline
\hline S6 & Seven Views + One & International Conference on Electrical and Information Technologies & E & Lahboube et al. & 2016 & \cite{Lahboube2016}  \tabularnewline
\hline S7 & On the Interplay of Business Process Modeling and Missions in Systems-of-Information Systems & International Workshop on Software Engineering for Systems-of-Systems & E & Graciano Neto et al. & 2017 & \cite{Neto2017} \tabularnewline
\hline S8 & Mandala: An Agent-Based Platform to Support Interoperability in Systems-of-Systems & International Workshop on Software Engineering for Systems-of-Systems & E & Mendes et al. & 2018 & \cite{Mendes2018} \tabularnewline
\hline S9 & Modeling Information Systems as Systems of Systems  & INCOSE Italia Conference on Systems Engineering & E & Salvaneschi & 2018 & \cite{Salvaneschi2018} \tabularnewline
\hline S10 & Representing Processes of Human Robot Collaboration & Workshop on Digitalization: Knowledge Work, Learning, and Industry 4.0 & E & Weichhart & 2018 &  \cite{Weichhart2018} \tabularnewline
\hline S11 & A Study on Goals Specification for Systems-of-Information Systems: Design Principles and a Conceptual Model & Brazilian Symposium on Information Systems & E & Graciano Neto et al. & 2018 & \cite{Neto2018} \tabularnewline
\hline
\end{tabular}
\end{center}
\end{table*}

\subsection{Overview of primary studies}
\label{sec:overview}

Table \ref{tab:primarystudies} shows that most studies (9 of 11) were published in the last four years (2018 was the latest year with a higher number of studies (4 of 11)). This trend shows an increasing recent interest for SoS BP, aligned to the era when organizations have tried to interplay their business processes and systems. 
Regarding the publication venues, most studies (9 of 11) were published in events (including 3 in workshops) and only 2 in journals, pointing out SoS BP as a novel topic with studies still not mature enough to be published in journals.
Studies are pulverized in seven different events and two journals, indicating that there are not still venues concentrating studies on SoS BP.

\begin{figure*}[h]
  	\centering
  	\includegraphics[width=120mm]{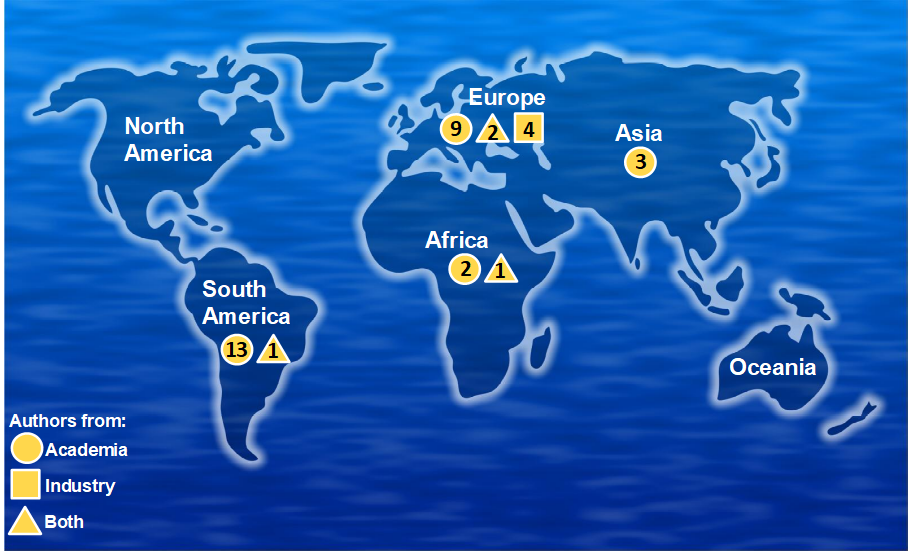}
  	\caption{Distribution of authors per origin/institution and per continent.}
  	\label{fig:authorDistribution}
  \end{figure*}

Most authors (27 out a total of 35) are solely from academia as noted in Figure \ref{fig:authorDistribution}, but there are authors from industry in most studies (7 of 11 studies), bringing some evidence of the importance of this topic for industry and also highlighting the need to apply research methodologies based on action research \cite{gustavsen2008}.
Moreover, the authors are distributed by four continents (South America, Europe, Africa, and Asia) as shown in Figure \ref{fig:authorDistribution} and in most continents, except one (i.e. Asia), there is collaboration between authors from both academia and industry, reinforcing the importance of this interaction for the advancement of research topic.

\begin{figure*}[h!]
  	\centering
  	\includegraphics[width=150mm]{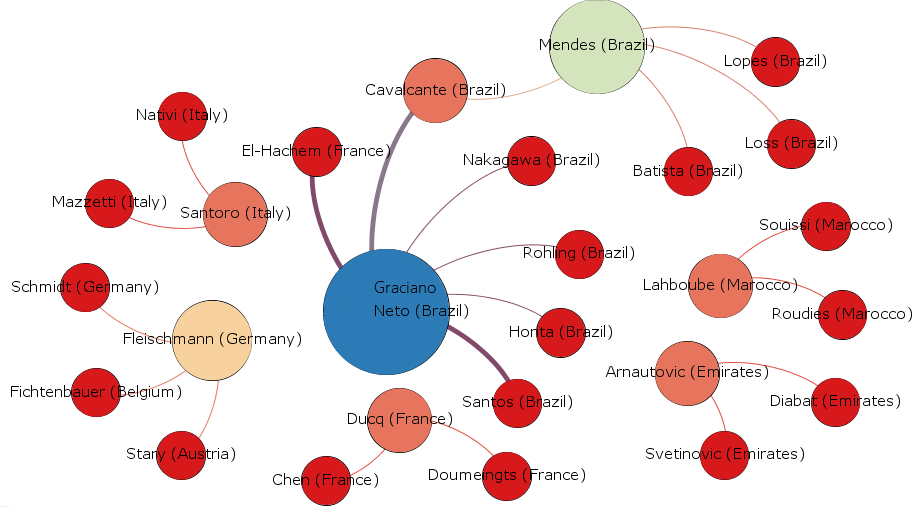}
  	\caption{Collaboration network among authors of primary studies.}
  	\label{fig:authorCollaboration}
  \end{figure*}

Another finding is that the research in SoS BP still occurs in isolation with only a punctual collaboration between research groups, as presented in Figure \ref{fig:authorCollaboration}\footnote{Generated using Gephi (\url{https://gephi.org})}, where the size of circles indicates the amount of relationship between the first author of each study and other authors. For instance, Santoro (from Italy) authored together with two authors, while Graciano Neto (from Brazil) authored with other six. Moreover, three authors do not appear in this figure, since they authored alone (S3, S9, and S10).
Most authors (19 of 27), in red circles, participated in only one study and are not first authors. These findings again confirm that the research in SoS BP is recent and further research collaboration should be consolidated.

\begin{figure*}[h!]
  	\centering
  	\includegraphics[width=250pt]{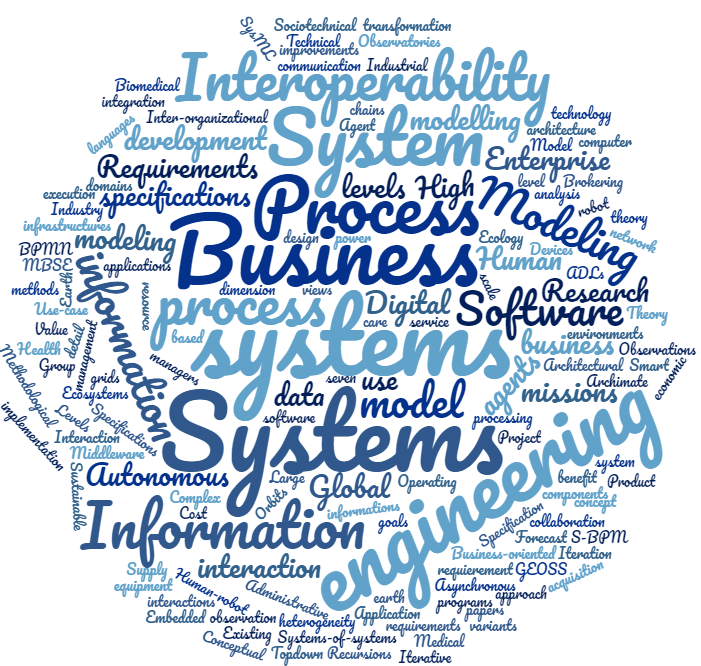}
  	\caption{Word cloud from keywords and abstracts of primary studies.}
    \label{fig:wordcloud}
  \end{figure*}

We also investigate which terms are addressed by the studies. For this, we generated a word cloud using the abstracts and keywords of all selected studies. As noted in Figure \ref{fig:wordcloud}\footnote{Generated using \url{https://www.wordclouds.com}}, besides more general terms (e.g., engineering, information, and software), studies address ``\textit{business process}'' and ``\textit{systems}'', as expected. Studies also mention ``\textit{interoperability}'', since it is a paramount issue in SoS, considering that the communication and information exchange among constituent systems could be addressed as early as possible (at the business process level), possibly spending less time and effort for the SoS development and evolution.
Another term is ``\textit{modeling}''; in fact, it is the first and usually the most important activity of BPM \cite{weske2019,van2013}.

\newpage
\vspace{-0.4cm}
\subsection{Rationales to deal with 
SoS BP}
\label{sec:discussion_rq1}

RQ1 identified the rationales (needs) raised by the studies to deal with the SoS BP and also the phases of the SoS life cycle that the SoS BP could support, as summarized in Table~\ref{fig:papersrationale}. 
%
We can observe the rationales 
are quite diverse.
The majority (S1 to S6, S8, and S10) build business process models to be used for the development of SoS. S7 and S11 started discussions directly related to SoS BP and defined broader guidelines for specifying missions of Systems-of-Information Systems (SoIS), a type of SoS where their constituents are information systems, without however pointing out how to do that.

 \begin{table}[h]
  	\centering
  	\caption{Rationales to Deal with SoS BP and Phases of the SoS Life Cycle where SoS BP have been used.}
  	\includegraphics[width=100mm]{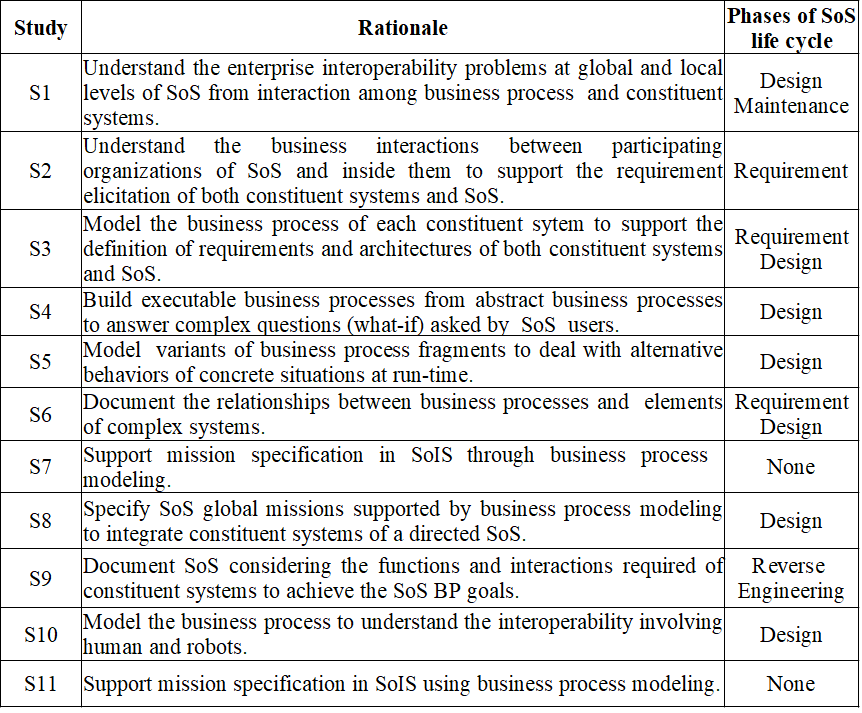}
  	\label{fig:papersrationale}
  \end{table}

S1 and S10 provided approaches that use business models to identify enterprise interoperability issues, while S2 and S3 presented approaches that build this model to support requirement elicitation (of both constituents 
and SoS) and, in addition, the approach supplied by S3 also used it to establish software architectures 
(for constituents 
and SoS). 
The approaches from S6 and S9 document SoS; 
while S6 mapped the SoS and its constituents to the corresponding business processes of involved organizations, 
S9 explained the constituents' elements to accomplish the SoS BP goals. Finally, S4 introduced an approach to build executable business processes from abstract business processes, S5 modeled variants of business process fragments to meet different needs of stakeholders, and S7, S8, and S11 dealt with SoS mission specification supported by process models.



With regard to the target phase of the SoS life cycle where the approaches 
could impact, some studies explicitly indicated it, but others do not. 
In the latter case, we presumed such phases (as listed in 
Table \ref{fig:papersrationale}) based on 
which phase the 
process model was used. 
Most studies addressed the earlier phases of the SoS life cycle, i.e., requirements and design/architecture, while only one study targeted maintenance. Other important phases for assuring the SoS quality like testing are still left out and, therefore, SoS BP needs still be better explored in the whole SoS 
life cycle.

In the state of the practice, SoS have been mostly built in an ad-hoc fashion, without 
considering larger business processes that could guide such building, 
e.g., providing requirements or supporting the architectural design. At the same time, constituent systems are usually developed without considering their further participation in a SoS
\cite{Dahmann2008}). 
Studies have initiated contributions in such directions: 
around half of the studies support requirements elicitation and two-thirds 
are concerned with SoS design.

\vspace{-0.3cm}
\subsection{SoS BP modeling}
\label{sec:discussion_rq2}

RQ2 discusses on the diverse means (methods,  modeling languages, techniques, and supporting tools) used by approaches to model SoS BP. Table~\ref{fig:instrumentsused} summarizes these results. We found only one approach that explicitly used a method to model SoS BP; moreover, different languages and techniques were adopted by the approaches, while only one approach pointed out the use of a supporting tool.

 \begin{table}
  	\centering
  	\caption{Means to Model SoS BP, SoS characteristics and activities of SoS BP life cycle and maturity level covered by approaches.}
  	\includegraphics[width=99mm]{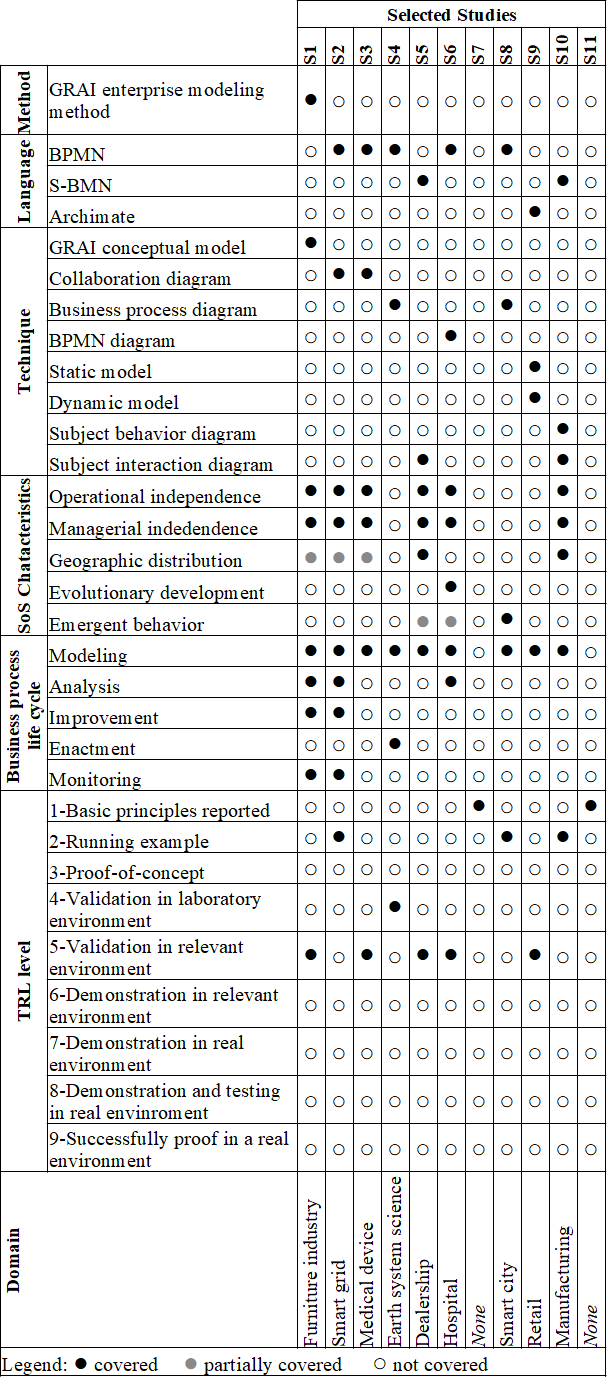}
  	\label{fig:instrumentsused}
  \end{table}

In detail, S1 applied a method for enterprise modeling called GRAI (Graph with Results and Activities Interrelated) \cite{doumeingts2001}, which is based on the system theory \cite{bertalanfy1968}. 
Firstly, this method creates a business process model (as-is model), highlighting the collaboration among organizations and the software of these organizations (called organizations’ network by authors) and identifying the enterprise interoperability issues. Then, the method interactively raises the strategic goals and gradually implements improvements to accomplish the enterprise interoperability.

Regarding the languages, our SR identified 
various diagrams, standing out those from BPMN (Business Process Model and Notation) \cite{omg2011} and S-BPM (Subject-oriented Business Process Management) \cite{Fleischmann2012}. 
Both 
are already peculiar for the business process specification,
and BPMN is a \textit{de facto} standard widely used. 
BPMN is not a formal language, but there is a mapping between BPMN and BPEL (Business Process Execution Language) \cite{omg2011}, which 
enacts business processes \cite{van2013} through services orchestration and deployment of the solution as a Web service; however, no approach found in our SR adopted BPEL. 

In turn, emerged after BPMN, S-BPM is an approach for specifying business processes centered on the acting elements in a process, i.e., the subjects (e.g., departments, people, robots, so on). Besides, S-BPM is based on a formal language that uses three basic elements (subject, predicated, and object) to enable stakeholders specify and execute business processes. Two studies (S5 and S10) adopted S-BPM in which the approach provided by S5 models variants of business process fragments and the approach from S10 models business processes involving human and robots, 
but they do not use this language (supported by tool) for simulation or enactment. Hence, due to S-BPM characteristics and mainly its formalism, \textit{it could be more widely adopted not only to model SoS BP but also to 
simulate and enact them}. 

With respect to the techniques, both collaboration diagram and business process diagram from BPMN and subject interaction diagram from S-BPM are the most used. Collaboration diagram can represent different business processes together
and the interaction among them through message flows, characterizing the choreography. Otherwise, business process diagram can represent only one process with its elements and the communication among them \cite{omg2011}, designating the orchestration. Subject interaction diagram can specify the interaction among subjects, such as people and systems, through messages containing business objects~\cite{Fleischmann2012}.
Another approach (presented in S6) only mentioned that a BPMN diagram was used, but did not specify which one. Other techniques were applied, but without expressiveness. 


With respect to supporting tools, they can reduce time and effort usually spent to model business processes
and contribute to the quality of these processes. Only one approach 
(in S9) mentioned the use of a tool (Archi tool\footnote{\url{https://www.archimatetool.com}}), which, although not directly destined to business process modeling, made it possible to create static and dynamic models of 
business processes.
These models can also show how each constituent system supports the SoS BP. 

Finally, all studies did not provide details about how the SoS BP modeling itself could be performed in a systematic way. \textit{More studies are then required to define how to step-to-step model SoS BP, considering also the inherent SoS characteristics}. 


\vspace{-0.4cm}
\subsection{SoS characteristics considered by the approaches}
\label{sec:discussion_rq3}

RQ3 sheds light on which SoS characteristics have been considered by the approaches, knowing that particularities of SoS can impact the business process level. 
We extracted the SoS characteristics when explicitly pointed out the selected studies. Observing Table \ref{fig:instrumentsused}, none approach covered all 
SoS characteristics.
Considering 
approaches that covered at least one characteristic, 
most (except from S8) contemplated ``operational independence'' and ``managerial independence'', where these approaches modeled each constituent organization having a distinct business process. Most of them (except S6 and S8) also attended 
``geographic distribution'', in which they clarified
the communication among the organizations' business processes of an alliance of organizations. Meanwhile the approaches from S1, S2, and S3, despite representing the communication among processes, they did not discuss how this communication should be managed. Hence, we classified them as partially covering ``geographic distribution'', as Table \ref{fig:instrumentsused} shows.
In more details, approaches that used the collaboration diagram (S2 and S3) and the subject interaction diagram (S5 and S10) modeled the business processes of different organizations, making it possible to explicit the exchange of messages among such processes and, consequently, represent their ``operational and managerial independence'' and the ``geographic distribution'' of these processes, which are directly impacted by the SoS inherent characteristics. Hence, SoS characteristics (i.e., operational and managerial independence and the geographic distribution) were considered in these approaches, aligned to the proper use of basic good practices for business process modeling, e.g., representing each role - organization or department or person - in a unique business process and the interaction among processes through message flow.

S6 presented an extended metamodel to design a multi-level view that can represent both elementary business processes and the macro, large business process. Hence, this approach covered  ``operational independence'' and ``managerial independence'', but it did not cover ``geographic distribution'' as this metamodel cannot explicitly represent the communication among these processes.
%
Only S6 considered the ``evolutionary development'', as this metamodel maps the changes that must be considered during the evolution of both constituents and SoS. 
With respect to ``emergent behavior'', only 
S8 attended it and two (S5 and S6) partially covered it. 
S8 modeled each SoS global mission using a business process diagram, representing SoS behaviors at business process level. S5 represented variants of business process fragments, which are 
realized at run-time. 
According to 
the authors, the replacement of variants is possible with minimal effect on other parts of the process, at both design-time and run-time, and such reconfiguration facilitates emergence of new behavior. The authors did not provide details about how this reconfiguration could be done and they did not mention how faults or unexpected behaviors in the business processes could be addressed. 
Finally, S6 identified the emergent behavior of SoS through the extended metamodel, but no means to accomplish the emergent behaviors was discussed. 

This scenario leads us to interpret the approaches are not still concerned with all SoS peculiarities at the business process level. More studies are need \textit{to properly include SoS characteristics during the SoS BP modeling, mainly that concerning the emergent behavior
, knowing that this characteristic can change the behavior of SoS BP at run-time and mitigating issues at business process level, such as interoperability 
and adaptability.}

\vspace{-0.2cm}
\subsection{Coverage of the SoS BP life cycle}
\label{sec:discussion_rq4}

RQ4 discusses which activities of SoS BP life cycle (modeling, analysis, improvement, enactment, and monitoring) were considered by the approaches, as shown in Table~\ref{fig:instrumentsused}. %
Noticeably, modeling is the most recurrent one (presented by 9 of 11 studies, S1 to S6 and S8 to S10). Other two studies (S7 and S11) presented more general guidelines to use business processes for SoS mission specifications, as mentioned earlier, not specifically discussing about modeling itself.

Regarding studies 
on modeling, two focused on using business process models to understand the organizational interoperability and interoperability among constituents of a SoS. S1 provided an approach that creates an as-is business model, highlighting the collaboration intra- and inter-organizations and their software systems.
%
S10 supplied an approach that builds business process models to represent the types of interaction between humans and robots.

With respect to studies that used process models to support requirement elicitation,
S2 creates a process model to represent economic entities/organizations and the corresponding systems into different types of entities, as well as the exchange of messages between them.  
%
S3 only mentioned 
the information concerning business processes 
is represented into a process model and optionally a description of the most relevant business processes can be elaborated that could be used to design the software architecture of constituents and SoS. 
Similarly, regarding studies that adopt process models to support SoS mission specification, S8 
builds business process models to represent activities associated 
with each constituent involved in each SoS global mission. The execution sequence of SoS's constituents is defined by sequence flows among the activities.

Studies also used process models to document SoS. S6 provided an metamodel to represent such models, which organize the knowledge on several processes, their descriptions, and a set of corresponding requirements of the SoS. 
S9 
created a static model in Archimate 
(that represents constituents
(components)
and 
relations among them
(connectors)) and a dynamic 
model also in Archimate (that represents the activities of such components and flows of data between activities) for each process of the SoS BP. 

The approach presented in S4 builds abstract business process models representing interactions among the business process elements, aiming at 
generating executable business processes.
Only one study addressed alternative behaviors of business processes. S5 proposed to model both the process fragments (containing involved subjects, their interaction, and their individual behaviors) and their variants, 
intending to attend different behaviors at run-time.

The analysis activity can assure the quality of the business process models (that can properly represent business goals and quality attributes, e.g., interoperability, recoverability, modifiability, and adaptability) at both design-time and run-time, but this activity was found in an unimpressive amount of selected studies and also in an incipient way. 
In more details, S1 provided an approach that compared
the as-is process and the ``target'' process (that corresponds to the strategic objectives of the alliance of organizations). 
S2 only mentioned that business process models could be analyzed by experts to check feasibility, complexity, and potential areas for improvement, e.g., via simulation, but without detailing how to do it. 
Lastly, S6 proposed to analyze and validate business process models by establishing a mapping between each business process and the SoS requirements. 

Similarly, improvement, enactment, and monitoring were still poorly investigated. %
In particular, the approach provided by S1 suggested to gradually improve the alliances' business processes to accomplish the enterprise interoperability and monitor its evolution through a performance indicator. 
S2 suggested to use business process mining techniques \cite{van2011} to both monitor and measure performance of the running business processes, but without detailing how to perform both.
Finally, S4 presented an approach to generate an executable business process (workflow) from an abstract business process and then this workflow (encoded in an executable language and supported by a tool) is uploaded into a workflow engine to be enacted.

We believe that, since modeling is usually the first activity of the business process life cycle and is responsible for building process models that are an essential artifact to start the BPM, for the while, most existing approaches have focused on this activity. The same trend occurs with the second activity that is analysis with three approaches, while improvement, enactment, and monitoring have not received much attention, yet. 
There are also approaches encompassing more than one activity, for instance, three studies (S1, S2, and S6) combined modeling and analysis, which are very close activities. When SoS BP topic is more mature, there will be certainly approaches covering many of these activities because all activities together should be considered for an effective management of SoS BP. \textit{Finally, we stand out the importance of investigating how to properly model, analyze, improve, enact, and monitor SoS BP to entirely accomplish the business goals of alliances of organizations supported by SoS.}

\vspace{-0.4cm}
\subsection{Maturity level of approaches 
}
\label{sec:discussion_rq5}

Considering that any new approach/solution should have a certain level of maturity to effectively contribute to a given area, RQ5 elucidates the maturity of the approaches.
To define the level of maturity, we inspired on the Technology Readiness Level (TRL), which allows 
to estimate the maturity level of not only 
technologies and/or systems on a scale from 1 (lowest level) to 9 (highest one) 
\cite{Mankins2009} but also software development processes with proper adaptations \cite{blanchette2010}. Thereby, we adjusted TRL to our work, as presented in Table \ref{fig:instrumentsused}.

Two approaches (in S7 and S11) are in the lowest level, since they only proposed general guidelines to specify SoS missions using business processes. 
Three approaches (in S2, S8, and S10) are in level 2 by presenting running examples.
One approach (S4) was classified in level 4, as it used concrete problems of three international research projects. 
%
Almost half of the approaches (5 of 11 studies) have level 5, i.e., they were evaluated in industrial environment;
however, these studies did not deep still the discussion. 
%
%
Hence, no approach was 
proven in depth in relevant and/or operational environments and
classified in levels 6 to 9.


Almost all approaches were used in a given application domain
(Table \ref{fig:instrumentsused}): some in critical domains (S2, S3, S4, and S8), industrial domains (S1 and S10), commerce domains (S5 and S9), and health (S6). 
%
%
These various applications reinforce the importance of looking at large business processes of the different domains with distinct complexity and levels of criticality.
Due to size and complexity of these domains, 
approaches were validated only using a part of the system/process. 
%
Except that provided by S4 that is very domain specific, approaches found in the studies could be 
explored in other domains; 
but their validity cannot be generalized yet. This fact also indicates the approaches are not completely mature to be applied in any SoS BP. 

\vspace{-0.3cm}
\section{Discussions
and Future Directions
}
\label{sec:discussions}

SoS BP are inherently complex and dynamic in the sense that they change at run-time due to external agents such as new laws, or internal issues, such as changes in the business goals of the organizations involved in the alliance or distinctive behavior of constituent systems (joining or leaving the SoS, being replaced, or changing or even failing). Thereby, these processes must properly deal with emergent and unexpected behaviors, but the \textit{existing approaches cannot still provide effective, mature support to holistically manage SoS BP}.
%
%
%
Considering this scenario, there are various future directions for research and practice. 
%
To the modeling of SoS BP, it is paramount to:
(i) understand which organizations can collaborate to achieve the business goals of alliances 
with greater profitability and technical excellence;
(ii) know how each organization can contribute to the business goals;
(iii) be aware of when and how should be the interaction among the business processes of involved organizations; 
(iv) understand how unforeseen issues, resulted from processes' interaction, can be handled; and
(v) define how to explicitly consider 
SoS characteristics as a whole during the 
SoS BP modeling.
Research from the area of networked enterprises \cite{gou2000,gou2003,meng2006,vanderhaeghen2007,akatkin2019} could be further explored. The issues raised from (i) to (iii), and partially in (v), could be addressed through a global view 
(which represents the entire networked enterprise's business processes containing general conditions for its member enterprises)
\cite{vanderhaeghen2007} 
and a local view 
(which refers to business process 
of each member
that meets public constraints 
recognized by all 
members) 
\cite{gou2003}. 
Still regarding 
(iii), 
interoperability solutions proposed 
for networked enterprises could 
be 
explored in SoS BP. 
For instance, one solution extended Petri Nets to address interoperability models 
from Workflow Management Coalition (WfMC)\footnote{\url{http://www.wfmc.org}} \cite{gou2000}, while 
another incorporated semantic annotation into executable BPMN models to achieve semantic interoperability \cite{akatkin2019}.

Techniques/languages 
used to model business processes 
are sufficient to deal with some SoS characteristics (in particular, the operational independence, managerial independence, and distribution of constituents), but 
not 
enough to cope with other peculiar characteristics 
mainly 
the dynamism that directly affect the SoS BP behavior. These languages are imperative (also known as procedural) and can explicitly specify all execution alternatives \cite{rodriguez2018b} by focusing on how the process work \cite{Fahland2009}. Otherwise, declarative languages (such as Declare) implicitly specify execution alternatives through constraints \cite{rodriguez2018b} prohibiting undesired behavior \cite{Haisjackl2016} (i.e., anything is possible unless explicitly forbidden \cite{van2013}) by emphasizing on the logic that governs the interplay of process actions \cite{Haisjackl2016}. In fact, declarative models provide a high degree of flexibility; however, problems to understand and maintain them, mainly due to hidden dependencies caused by the combination of constraints, often hinder their adoption \cite{Haisjackl2016}. Hence, a possible solution to soften these drawbacks is to harmoniously combine imperative and declarative languages. 

Late modeling, which consists of specifying parts of the process or the entire process at run-time \cite{Weber2009}, could be adopted to SoS BP 
can foresee emergent behaviors at design-time and 
perform the necessary, adequate configurations at run-time.
%
%
Alternatively, such dynamic process could be modeled using ad hoc sub-processes from BPMN (i.e., set of activities without a defined sequence at design-time \cite{omg2011}) associated with scenario descriptions 
(which specify parallel activities and sequence of activities) \cite{Szelagowski2019b}. Another initiative 
is the dynamic workflow patterns \cite{Lam2015}, which can determine at run-time the processes elements and 
the communication among them. 
Besides, initiatives from the networked enterprises area could shed light on the dynamism of SoS BP. For instance, we could explore agent-based approach that yields the description of dynamic control structures of networked enterprise’s business processes \cite{gou2003} or could propose an extension of underlying models in Workflow Process Definition Language (WPDL) to specify dynamic properties of inter-organizational process models \cite{meng2006}.
Hence, we believe research on dynamic business processes, including those in the context of networked enterprises, could be revisited and experimented with in SoS BP.

Considering the advantages of formal languages for business process specification, such as YAWL \cite{Hofstede2010} and S-BPM, besides their well-defined syntax and semantics and the ability to support completeness, consistency, and correctness analysis, it seems to be interesting to adopt them to properly model and also simulate SoS BP.
%
%
%
%
%
To the \textit{analysis of SoS BP}, it is important to assure that these processes are correct concerning syntax, structure, and behavior (related to functional and non-functional business requirements), guaranteeing correctness on:
    (i) interaction among business processes of distinct organizations;
    (ii) behaviors resulting from this interaction; and
    (iii) handling of possible failures resulting from this interaction.
Quantitative techniques, such as cost-related measures, time-related measures, and quality-related measures that are recurrent classes of measures in the context of BPM \cite{dumas2018}, can be adapted to the analysis of business processes of each organization together with the SoS BP itself.

Regarding the \textit{SoS BP improvement}, it is 
first necessary to identify how to deal with changes in these processes previously raised in the analysis activity. Soon after, the most appropriate changes must be done using appropriate process analysis techniques \cite{dumas2018} to ultimately meet the business goals of alliances of organizations.
To the \textit{enactment of SoS BP}, as also indicated to the traditional BPM, it is necessary to address changes in the organizational management and process automation \cite{weske2019, van2013, dumas2018}, but adapted to the particular case of SoS BP. In a nutshell, that implies changes in how all stakeholders involved in the SoS BP work, as well as development/evolution and deployment of software systems (i.e., constituents), which compound the SoS and, in turn, support the SoS BP. 
In addition, experience from the enactment of networked enterprises’ business processes (which are mainly concerned with collaboration and interoperation of business processes of member enterprises \cite{gou2003,meng2006,baina2006}, for instance, using web services \cite{meng2006,li2010} and ``business services'' (external services to an enterprise to execute its specific process activities) \cite{baina2006}) could be investigated and possibly extended to SoS BP.
Regarding the \textit{SoS BP monitoring}, adapted from the original BPM \cite{weske2019, van2013, dumas2018}, relevant data about this large process and, as a consequence, data from the business processes of each organization must be collected and together analyzed at run-time, aiming at observing whether all of them are working in accordance with performance measures and accomplishing the large business goals. However, this task is not trivial because the processes involved in the alliance have multiple owners and stakeholders. Monitoring of SoS BP will also make it possible to identify new business demands/circumstances and emerging technologies that must be also treated. 




Undoubtedly, supporting tools must be adopted to manage SoS BP. Well-known tools, such as the commercial ones like Bizagi\footnote{\url{https://www.bizagi.com}}, BPMN2 Modeler\footnote{\url{https://www.eclipse.org/bpmn2-modeler}}, and BPM \textit{inspire}\footnote{\url{https://marketplace.eclipse.org/content/bpm-inspire-0}}, as well as academic tools like YAWL
\footnote{\url{https://yawlfoundation.github.io}} and KIT S-BPM Modeler\footnote{\url{https://github.com/mkolodiy/s-bpm-modeler}}, 
provide functionalities of modeling and some of them 
also enable analysis 
(e.g., Bizagi and YAWL) and enactment
(e.g., a dynamic workflow management system   \cite{meng2006} and a business process-oriented heterogeneous systems integration platform \cite{li2010}).
All of them support the use of good practices of business process modeling (e.g., those recommended as the workflow patterns\footnote{\url{https://www.workflowpatterns.com/}} \cite{Russell2016}, such as control-flow-oriented workflow patterns, workflow resource patterns, workflow data patterns, and exception handling patterns), which are also necessary for SoS BP. 

As available nowadays, these existing tools can model processes that comply with three SoS characteristics (operational independence, managerial independence, and distribution of constituents), because it is possible to represent each role -- organization, department, or person -- as a unique business process, to represent the interaction among processes through message flow, and to describe how this communication must be managed. These tools also provide exception handling, which can treat faults during the process execution. 
Meanwhile, none of them alone provide mechanisms to deal with SoS BP dynamism; for this, tools that implement variability techniques for business processes, such as YAWL tool combined with Synergia toolset \cite{Rosa2009} and/or BPL-Framework \cite{Ferreira2017a}, could be used. For the while, there no exists a tool that can automate all BPM activities in the adequate way to the SoS BP that, as stated before, are completely dynamic at run-time and directly affected by SoS inherent characteristics.


In summary, to make feasible future research directions on SoS BP (which could be considered as \textit{Systems-of-Systems meta-architecture} in other research areas), we suggest as immediate directions: 
(i) systematically revisit means (methods, languages, techniques, and tools) traditionally used in the BPM and networked enterprises areas, so that these means can be possibly adapted to SoS BP; 
(ii) develop appropriate means to manage SoS BP and, due to the diversity of techniques necessary for managing them, it is primordial to establish the way to the integration among them together with supporting tools;  
(iii) investigate how to assure quality attributes of SoS BP, such as interoperability, performance, scalability, and reliability, as well as metrics to measure them; and
(iv) conduct empirical studies in real-world SoS BP to observe the applicability of these means.


As main benefits of an appropriate SoS BP management, we presume that: 
(i) alliances of organizations can have a holistic comprehension of their SoS BP and their goals; and
(ii) each organization can comprehend its role and responsibility to accomplish the SoS BP goals. 
As a consequence, JV and M\&A will be more likely to succeed, differently of nowadays in which most of them fail due to diverse factors, 
including the lack of understanding of organizations' business processes involved in their large processes
\cite{kumar2019}. 
Another important benefits 
is the support to the SoS development and evolution, which are usually performed without taking advantages of information from business level that SoS BP can aggregate. More specifically, the identification of SoS missions and elicitation of requirements (to be implemented in the SoS and its constituents to attend large business goals/needs) can use information from business level contained in the SoS BP models; hence, dynamic business process is also possible to clarify the relationship between the business level (i.e., the SoS BP) and technical level (i.e., the SoS itself and its constituents). Last but not the least, sustainability and longevity of both the alliances of organizations and the SoS could be achieved in the sense that both can survive for a long time adapting smoothly at run-time to diverse circumstances.

\vspace{-0.3cm}
\section{Conclusion}
\label{sec:conclusion}

In the era of JV and M\&A, BPM becomes even more important due to large, complex, and interconnected business processes resulted from these new ecosystems formed of several distinct, distributed organizations and the SoS. In this scenario, the main contribution of this paper was to present a landscape of approaches being proposed to manage such large processes and, notably, we observe there is still a long way to mature such approaches and introduce them effectively in the context of alliances of organizations. Another intention of this work is to call the attention of research community and practitioners about the 
need of more research in the intersection of BPM and SoS and, more importantly, this work also intends to alert companies/organizations about the need of investment in this very new research topic, the SoS BP, which could be decisive to the success or failure of alliances of organizations.

For promoting SoS BP, besides many future research challenges 
to be overcome, real-world scenarios where SoS BP emerge should be the objects of investigation to learn from such scenarios, collecting experience and practice to support the consolidation of the area of SoS BP.

\section*{Funding}
This work was supported by Federal University of Mato Grosso do Sul and Brazilian funding agencies CAPES (Finance Code 001); FAPESP (grants: 2015/24144-7); and CNPq (grants: 313245/2021-5).

\bibliography{references}

\end{document}